\documentclass[11pt]{article}
\usepackage{graphicx}
\usepackage{lineno,hyperref}
\modulolinenumbers[5]
\begin{document}

\title{Chaos-based Potentials in the One-dimensional Tight-binding Model Probed by the Inverse Participation Ratio}


\author{
WESLLEY FLORENTINO DE \ OLIVEIRA\footnote{Permanent address: Departamento de Ensino, IFNMG - Instituto Federal do Norte de Minas Gerais, 
Pirapora, MG, Brazil.}\\
Programa de P{\'o}s-gradua\c c\~ao em Modelagem Matem\'atica e Computacional\\
CEFET-MG - Centro Federal de Educa\c c\~ao Tecnol\'ogica de Minas Gerais\\ 
Belo Horizonte, MG, Brazil. \\ \ \\
\and 
GIANCARLO QUEIROZ \ PELLEGRINO\footnote{Corresponding author (ORCID iD: 0000-0002-2555-2309).}\\
Departamento de Matem\'atica\\ CEFET-MG - Centro Federal de Educa\c c\~ao Tecnol\'ogica de Minas Gerais\\ Belo Horizonte, MG, Brazil.\\ \ \\
E-mails: weslley.florentino@ifnmg.edu.br \ \ giancarloqpellegrino@cefetmg.br
}

\maketitle

\begin{abstract}
Chaos-based potentials are defined and implemented in the one-dimensional tight-binding model as a way of simulating disorder-controlled crystalline lattices. In this setting, disorder is handled with the aid of the chaoticity parameter. The inverse participation ratio $(IPR)$ probes the response of the system to three different such potentials and shows consistent agreement with results given by the Lyapunov exponent $Ly$: the greater $Ly(r)$ for the chaotic sequence as a function of the chaoticity parameter $r$, the greater the asymptotic value $IPR(r)$ for the large-system ground state.
\end{abstract}


\noindent
{\bf Key words:} aperiodic potentials, tight-binding model, inverse participation ratio, Lyapunov exponent.


\section{Introduction}\label{sec:1}

Growing crystalline structures and new materials which would show previously desired properties --- or avoid undesired ones --- is a long-standing task. As a particular characteristic, disorder is an inherent and mostly uncontrolled feature of natural materials. It is acceptedly one of the main responsible facts for resistivity, be it electrical, thermal, or signal resistivity; and its very notion is based on the randomness of its occurrence. Contrary to common sense, controlled disorder can be made an interesting property, although perhaps difficult to achieve. The advent of quasicrystals made a notable step towards disorder control, due to the possibility of using binary substitution sequences to guide deposition-based crystalline growth. Although disorder was not fully controlled, it could be classified as an ascending feature for different quasicrystals, i.e., different substitution sequences. Remarkably, the early and well-known chaotic maps can give rise to binary sequences where the disorder is controlled by their chaoticity parameter and even measured by their associated Lyapunov exponent. Here, such chaos-based sequences are constructed from three known maps --- logistic, tent, and Gaussian maps --- and used as potentials in the one-dimensional Schr\"odinger equation. It is seen that a system simulated in this way not only responds unambiguously to more or less disorder in the potential but also allows for a disorder control in an experimental realization.

The remainder of this article is organized as follows.
Section~\ref{sec:2} states the main ideas, scope, and objectives of this work. In section~\ref{sec:3}, there appear the necessary definitions and model. Section~\ref{sec:4} brings results and comments. In section~\ref{sec:5}, we discuss the consistency of our proposals in comparison with results known for disordered potentials. A short section~\ref{sec:6} concludes this article.

\section{Theory}\label{sec:2}

Aperiodic sequences of different types have recently been used to simulate disordered one-dimensional crystalline structures, as potentials entering the Schr\"odinger equation. One of the main purposes of this effort is to simulate the physics of quasicrystals, which can be viewed as intermediate systems between the well-understood cases of periodic and disordered lattices. The latter ones are fairly well described by Bloch theorem and Anderson localization phenomenon respectively. In the case of periodic potentials, the eigensolutions of the Schr\"odinger operator can be chosen to exhibit periodic absolute values \cite{ashcroft}, thus providing good conducting properties. At the other extreme, one-dimensional disordered potentials afford localized probability densities and, in this way, insulating properties \cite{anderson1958}.

In the context of one-dimensional systems, quasicrystal issues have been addressed by using almost-periodic substitution sequences to mimic the ordering of the atoms in a crystalline lattice \cite{divincenzo1991,axel1995}. Substitution sequences and the properties of the Scr\"odinger operator have been extensively studied both from the theoretical \cite{queffelec1987,luck1989,dulea1992,ryu1992,dulea1993,oh1993,piechon1995,cesar1999} and the experimental \cite{merlin1985,todd1986,axel1991,mizoguchi1997} points of view. Some examples of these sequences are the early Fibonacci substitution sequence, as well as Thue-Morse, period-doubling, paper-folding, Rudin-Shapiro sequences, and others \cite{axel1995}. The main interests in these studies are the physical transport properties \cite{dulea1992a,iochum1992,ryu1993,roy1994,katsanos1995,roy1995,piechon1996,roche1997,salejda1998,pellegrino2001} and the mathematical spectral types \cite{queffelec1987,bovier1995,allouche1997} of the corresponding Schr\"odinger operators.

The almost-periodic sequences cited above have no definite period and appear as intermediate cases between periodic and disordered structures. Although these substitution sequences could be classified in a hierarchy of ascending disorder \cite{pellegrino2001,steuer2001,gong2015}, this is a static classification, since each sequence is generated by a fixed iteration rule. It would be greatly desirable both for the physical and mathematical possibilities to have a variable disorder, eventually controlled by a continuous variation of a parameter. This is the main purpose of this article. We start exploring here the properties of the Schr\"odinger operator triggered by a new type of possibly disorder-controlled potential, which will be based on iteration maps appearing in Chaos Theory \cite{hilborn} (hereafter called chaotic potentials)\footnote{To the best of the authors' knowledge, this was first proposed by C. R. de Oliveira (personal communication, 2002).}. As it is well known those maps are not periodic in the chaotic region, and chaoticity can be controlled by a continuous parameter $r$ and measured by a Lyapunov exponent \cite{hilborn}.

To that end, our main probing tool will be the Inverse Participation Ratio ($IPR$) of a given eigenvector of the Schr\"odinger operator. This ratio is a real number defined and used in Solid State Theory to characterize localization-delocalization transitions and their consequences to the transport properties \cite{evers2000,ludlam2004,monthus2010,murphy2011}. It will be seen that $IPR$ for the ground-state eigenvector, as a function of the chaoticity parameter, shows very good agreement with the structure of chaotic regions and periodicity windows characteristic of these chaotic maps. In this way, one could seek to control the effects of the disorder in the potential sequence appearing in the Schr\"odinger operator just by varying the chaoticity parameter.

\section{Definitions and model}\label{sec:3}

\subsection{The tight-binding model}

We consider here the time-independent Schr\"odinger operator in the tight-binding approximation
\begin{equation}
\left( H\psi \right)_n = \psi_{n+1} + \psi_{n-1} + V_n\psi_{n}, \label{schreq}
\end{equation}
where one assumes the potential $V$, created by a lattice of atoms placed at positions $n$, to be concentrated in the regions close to the atoms, being negligible in the regions between them \cite{ashcroft,cesar1999}. In this approximation, the eigenfunctions are given by the values $\psi_n$ at sites $n$, when subjected to a potential with values $V_n$. In this model, one easily switches different potentials just by choosing appropriate sequences of real numbers $\left\{ V_n\right\}_{n=1}^{\infty}$.

\subsection{Chaotic sequences and potentials}

The sequences to be used as potentials in eq. (\ref{schreq}) will be obtained from three iteration maps $w_{n+1} = f_{r}\left( w_{n} \right)$ commonly encountered in Chaos Theory \cite{hilborn}, namely
\begin{itemize}
\item the logistic map
\begin{equation}
w_{n+1} = rw_{n}\left( 1 - w_{n} \right) \hspace*{1cm} r \in \left[ 3.4,4.0 \right] \label{logmap}
\end{equation}
\item the tent map
\begin{equation}
w_{n+1} = r\left( 1 - 2\left| {w_{n} - \frac{1}{2}} \right| \right) \hspace*{1cm} r \in \left[ 0.4,1.0 \right] \label{tentmap}
\end{equation}
\item the Gaussian map
\begin{equation}
w_{n+1} = e^{-b {w_{n}}^{2}} + r  \hspace*{1cm} r \in \left[ -1,0.1 \right] \label{gaussmap}
\end{equation}
\end{itemize}

In all these maps, $r$ is a continuous parameter which controls in a sense the chaoticity of the map. In the expressions above, the intervals are chosen to cover the regions of chaos and periodicity of the maps. For the sake of the point we want to make, plots for these maps will appear in the next section along with our results. It suffices to state here that, for $r$ in the chaotic regions, these sequences neither have definite period nor are completely random, once they are constructed with fixed rules. Chaoticity, which will give us the desired disorder, is incremented by going deeper into the chaotic region and is measured by the corresponding Lyapunov exponent \cite{hilborn}
\begin{equation}
Ly \left( r \right) = \frac{1}{N}\sum\limits_{n = 1}^N {\ln \left| {{f'_r}\left( {{w_n}} \right)} \right|}. \label{lyapunov}
\end{equation}

In order to allow an easier comparison with results found in the literature for almost-periodic substitution potentials, and also to set an experimental perspective based on crystal growing \cite{axel1991}, we take here binary chaotic potentials as given by
\begin{itemize}
\item $V_{n}=1$ if $w_{n} \leq 0.5$ and $V_{n}=2$ if $w_{n} > 0.5$ in the cases of logistic and tent maps;
\item $V_{n}=1$ if $w_{n} \leq 0$ and $V_{n}=2$ if $w_{n} > 0$ in the case of the Gaussian map.
\end{itemize}

The Lyapunov exponent, as given by eq. (\ref{lyapunov}), is not directly applicable to these binary sequences \cite{steuer2001,gong2015}. It is, however, remarkable the fact that the Inverse Participation Ratio, to be defined in the following, captures indeed the features of that exponent even after the transposition to the corresponding binary sequence. Details of this accordance will be seen in the next section.

\subsection{Inverse Participation Ratio}

Perhaps the most conspicuous effect of a one-dimensional random potential is the localization of the wavefunctions, solutions of the Schr\"odinger equation, within finite regions in space \cite{anderson1958,kramer1993}. At the other extreme, periodic potentials give rise to wavefunctions which are extended over the lattice \cite{ashcroft}. The Inverse Participation Ratio of a given wavefunction, defined as a function of the system size $N$ by
\begin{equation}
IPR\left( N \right) = \left[ \frac{\sum\limits_{n = 1}^N \left| \psi_{n} \right|^{4}}{\left( \sum\limits_{n = 1}^N \left| \psi_{n} \right|^{2} \right)^{2}}\right]^{-1},
\end{equation}
reflects the regions where the wavefunction has appreciable amplitudes and, therefore, its distribution over the lattice \cite{evers2000,ludlam2004,monthus2010,murphy2011}. In this way, one can verify that $IPR(N)$ tends to zero as $1/N$ for large enough system sizes $N$ in the case of periodic potentials. For the random case, $IPR(N)$ attends a constant value $k>0$ as $N$ goes to infinity \cite{ludlam2004,monthus2010}. Aperiodic potentials, such as the almost-periodic substitution potentials, are expected to fall between these extreme cases \cite{weslley2014}. The large-$N$ values of $IPR(N)$ could, therefore, probe the effect of a greater or smaller disorder in the potentials; the higher the asymptotic value of $IPR(N)$, the greater the disorder of the potential. Here we take the eigenvector associated with the lowest energy eigenvalue for the Schr\"odinger operator and calculate its large-$N$ value for $IPR$. This asymptotic value, being a function of the parameter $r$, will be denoted here simply $IPR(r)$. This will be our probing tool in order to see whether chaotic potentials, as defined here, could provide a way in which disorder is implemented in a more controlled procedure, as compared to other known aperiodic potentials.

\section{Results}\label{sec:4}

In Figures 1--3 we show in plots (a) the original chaotic maps for different values of $r$, prior to the conversion into binary sequences. In plots (b) there appear the Lyapunov exponents for the original sequences and the average asymptotic large-$N$ values $IPR(r)$ for the corresponding binary sequences. In all these plots, we use sequences of size $N = 1000$ and in the calculation of $IPR(r)$ we take averages over the results for $500$ realizations of the sequences with different initial seeds $w_0$.

\begin{figure}[h!]
\begin{center}                                     
 \includegraphics[height=10cm, width=12cm]{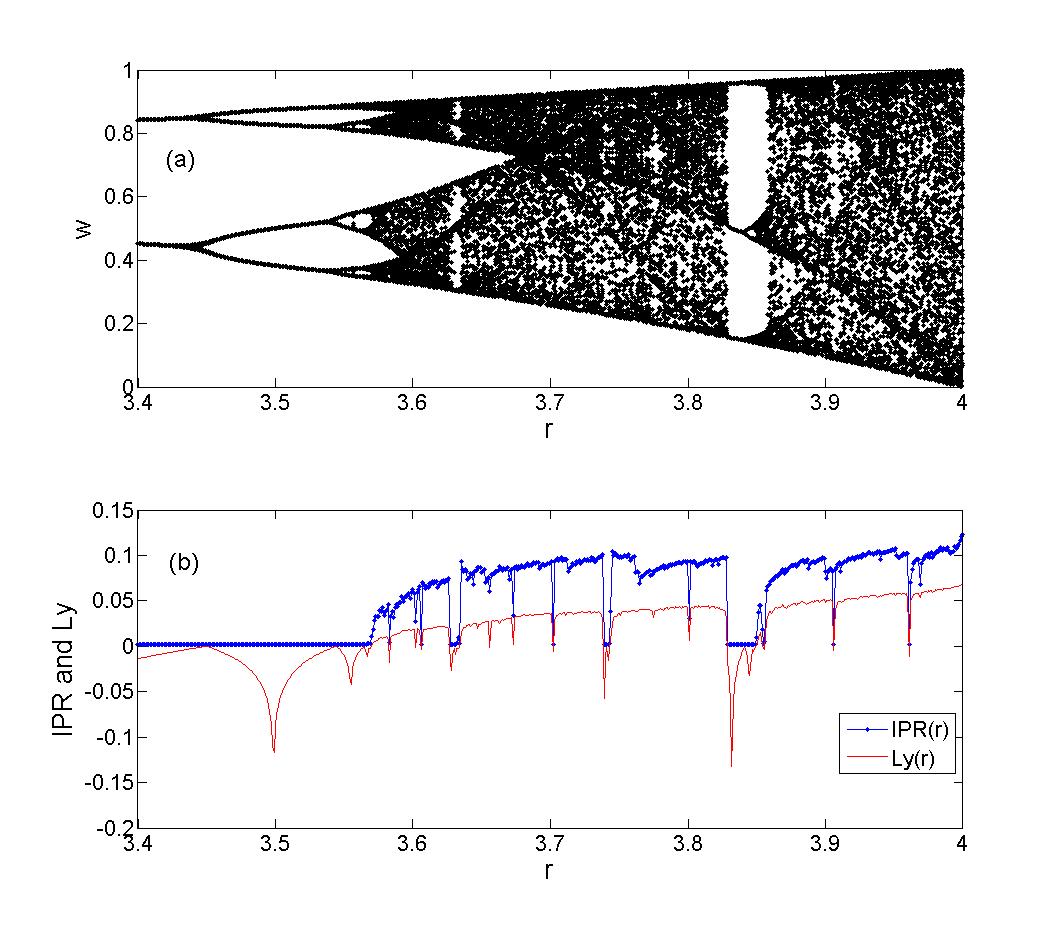} 
 \caption{Logistic Map. (a) Bifurcation Diagram for $r \in [3.4,4.0]$. (b) $IPR(r)$ and $Ly(r)$ multiplied by $0.1$ for better viewing.}
 \label{fig1}
 \end{center}
\end{figure}

\begin{figure}[h!]
\begin{center} 
 \includegraphics[height=10cm, width=12cm]{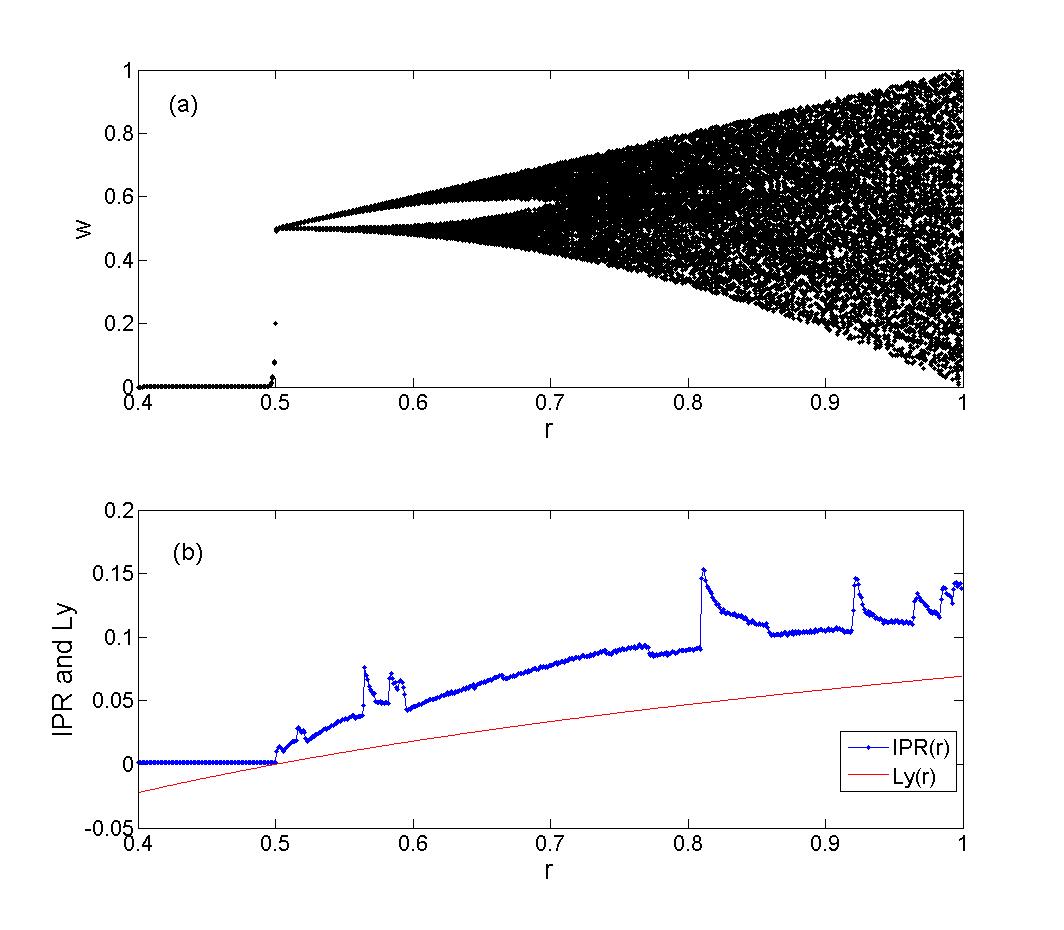} 
 \caption{Tent Map. (a) Bifurcation Diagram for $r \in [0.4,1.0]$. (b) $IPR(r)$ and $Ly(r)$ multiplied by $0.1$ for better viewing.}
 \label{fig2}
 \end{center}
\end{figure}

\begin{figure}[h!]
\begin{center} 
 \includegraphics[height=10cm, width=12cm]{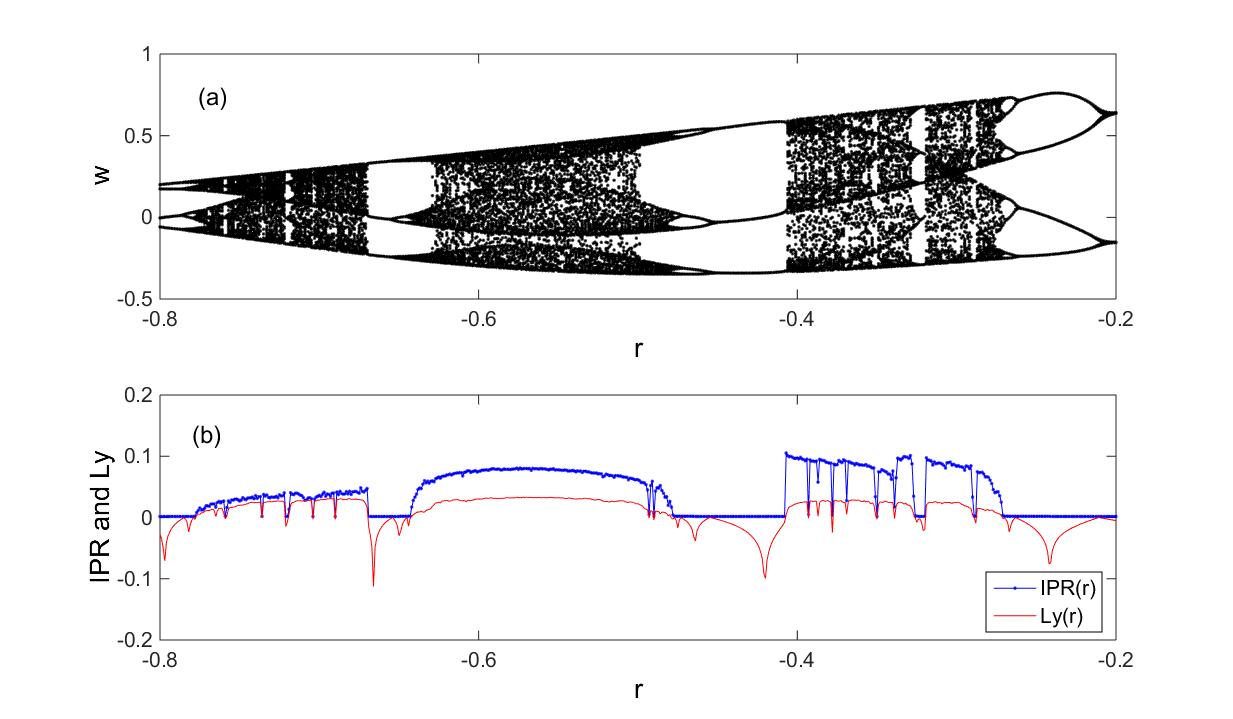} 
 \caption{Gaussian Map. (a) Bifurcation Diagram for $r \in [-1.0,-0.1]$. (b) $IPR(r)$ and $Ly(r)$ multiplied by $0.1$ for better viewing.}
 \label{fig3}
 \end{center}
\end{figure}

In plots (a) we see the characteristic features of these chaotic maps, with regions of chaos and, eventually, windows of periodicity. We have chosen these three maps due to their different routes to chaos. The logistic map undergoes a sequence of period-doublings for its fixed points until entering a region of chaos at $r \approx 3.6$. This chaotic behavior is observed up to $r = 4.0$, with periodicity windows appearing at precise values of $r$. For the tent map, the behavior is quite different, the sequence entering directly the chaotic region without passing a period-doubling transition. Finally, for the Gaussian map, we see the period-doubling transition to chaos, as in the case of the logistic map, but also a period-undoubling transition back to periodicity. The corresponding plots in (b) show that $IPR(r)$ accompanies this structure, with positive values for $r$ in the chaotic regions and vanishing values for $r$ in the periodicity windows.

This agreement is in fact remarkably emphasized by the comparison between $IPR(r)$ and the Lyapunov exponent $Ly(r)$, both shown in plots (b). Expression (\ref{lyapunov}) yields corresponding forms for the Lyapunov exponents associated with the original maps studied here (eqs. \ref{logmap}--\ref{gaussmap}):
\begin{itemize}
\item for the logistic map
\begin{equation}
Ly \left( r \right) = \frac{1}{N}\sum\limits_{n = 1}^N {\ln \left| {r\left( {1 - 2w_n } \right)} \right|}
\end{equation}
\item for the tent map
\begin{equation}
Ly \left( r \right) = {\ln \left| 2r \right|} 
\end{equation}
\item for the Gaussian map
\begin{equation}
Ly \left( r \right) = \frac{1}{N}\sum\limits_{n = 1}^N {\ln \left| {2bw_n e^{ - bw_n ^2 } } \right|}
\end{equation}
\end{itemize}

 In the regions where $Ly(r) \le 0$, $IPR(r)$ vanishes as well, pointing out a periodicity window. On the other hand, $Ly(r)$ is positive in the regions of chaotic behavior, a feature captured by $IPR(r)$. Moreover, one can see that $IPR(r)$ follows on average the increasing-decreasing behavior of $Ly(r)$. A closer look reveals that $IPR(r)$ increases smoothly as the maps undergo the bifurcation process, and drops down abruptly when the maps enter a periodicity window. This behavior can be seen, for example, at $r \approx 3.83$ for the logistic map, or at $r \approx -0.4$ for the Gaussian map. On the other hand, one observes some abrupt jumps in the $IPR$ curve for the tent map, which are apparently not associated with a particular behavior of the corresponding map. The wavefunction for these values of the chaoticity parameter, however, shows accordingly a more localized distribution. This is suggestive of a consistent estimative, via $IPR(r)$, of the response of the system to the disorder of the potential. In references \cite{pellegrino2001, gong2015} it is pointed out that different approaches in order to produce a hierarchy of disorder for aperiodic sequences frequently end at inconsistent results.

We stress the fact that this consistent agreement between $Ly(r)$ and $IPR(r)$ is seen here despite the conversion of the original sequences into binary ones, $Ly(r)$ being directly calculated for the former and $IPR(r)$ being indirectly calculated for the ground state generated by the latter ones. This agreement is found also for the excited states, with the same overall behavior of $IPR$. In the next section we give more details on this point.

\section{Discussion}\label{sec:5}

Much work on disordered potentials in one-dimensional tight-binding models has been done in the last decades. Since the potentials proposed here are less studied, we compare our results with those ones known for disordered potentials, in order to check for the consistency of $IPR$ as indicative of the effects of randomness.

The correlation between the ratio $IPR$ and the Lyapunov exponent $Ly$ of the sequences which generate these chaotic potentials is clearly seen in Figure 4, where we plot the varying of $IPR(r)$ with $Ly(r)$. In fact, wherever $Ly > 0$ one has $IPR > 0$. Also, $IPR$ increases on average with the increase of $Ly$. For logistic and Gaussian potentials, the many-valued character of $IPR$ is due to the periodicity windows in the original maps, where $Ly$ goes back to negative values and reincreases with the next period-doubling process, followed by $IPR$ which goes back to zero and reincreases jointly. In this respect, the tent map shows a neat single-valued aspect, since there are no periodicity windows above the chaos threshold. In a situation where one would seek controlled disordered, the corresponding tent potential could be a good choice.

\begin{figure}[h!]
\begin{center} 
 \includegraphics[height=10cm, width=12cm]{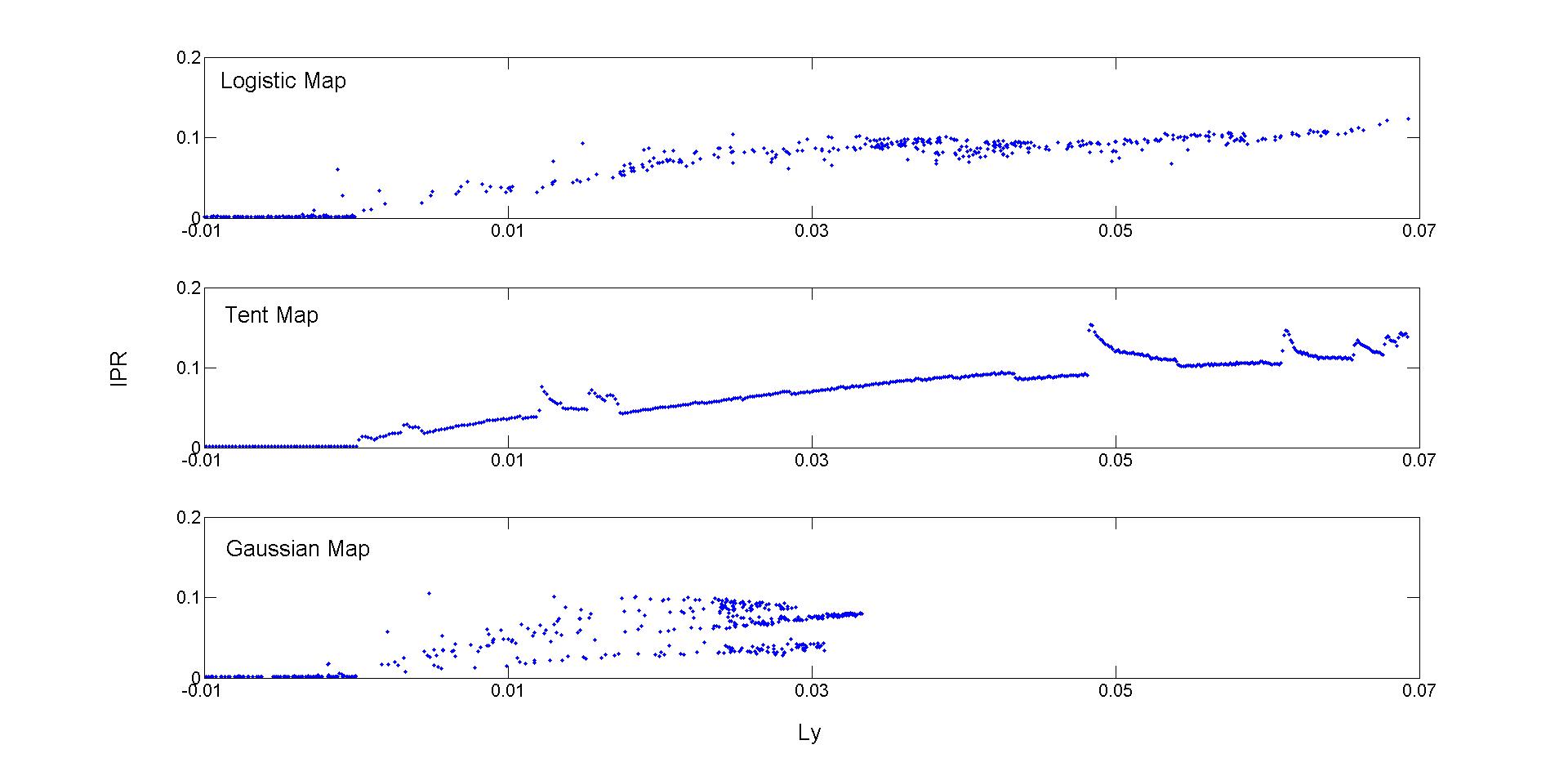} 
 \caption{Correlation between $Ly$ and $IPR$, both obtained as functions of the chaoticity parameter $r$.}
 \label{fig4}
 \end{center}
\end{figure}

For potentials with increasing randomness, the energy-band structure seen for periodic potential is gradually destroyed. It is known that, in this process, the eigenstates associated with band-edge energies are more localized than those associated with band-center energies. This delocalization feature is also captured by the inverse participation ratio, eigenstates with greater $IPR$ being more localized. We verified that this agreement is found for the excited states all along the spectrum, and we illustrate it with the logistic potential in the chaotic region ($r = 4$). In Fig. 5 we show the inverse participation ratio $IPR$ for all the eigenstates. The central region of the spectrum have on average eigenstates with lower values for $IPR$. From these results we choose three eigenstates ($k = 101, 510, 511$), and plot in Fig. 6 the corresponding probability densities ${\left | \psi_k \right |}^2$ over the lattice. It is seen that even small variations of $IPR$ indicate corresponding variations in the localization length of states.

\begin{figure}[h!]
\begin{center} 
 \includegraphics[height=10cm, width=12cm]{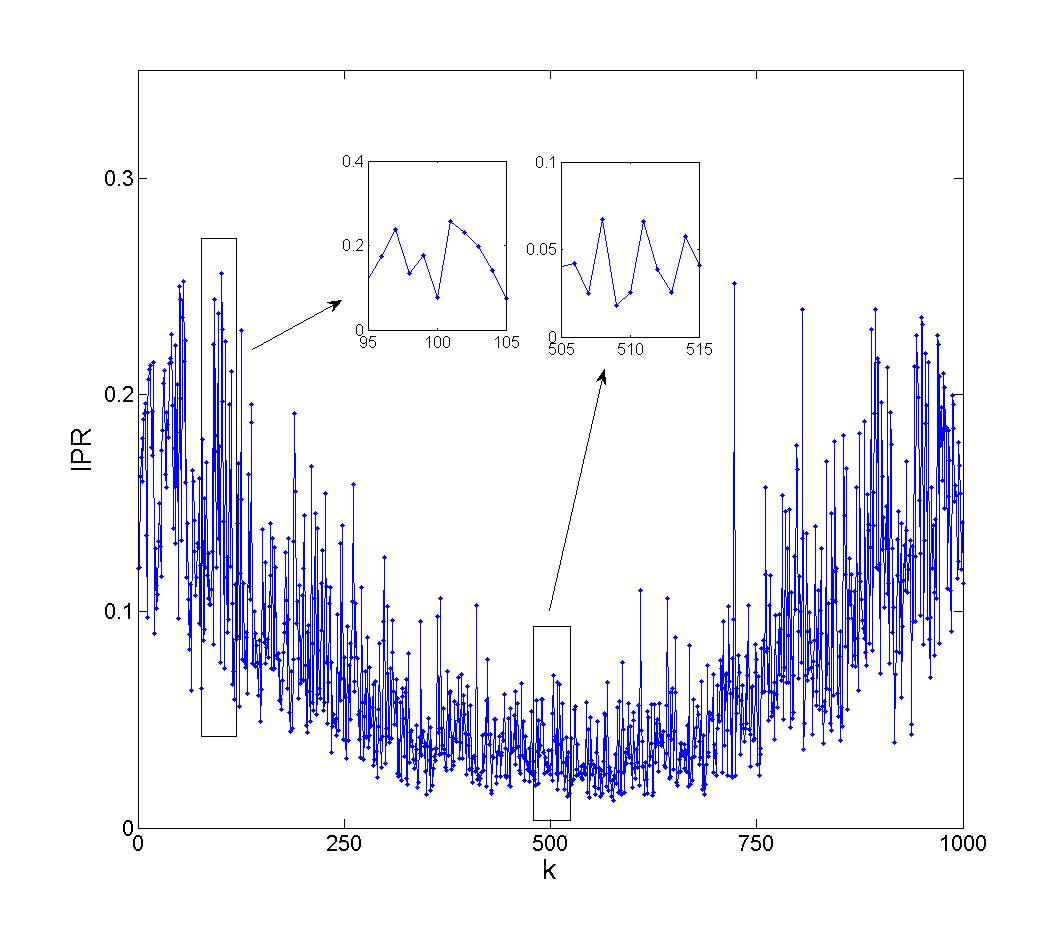} 
 \caption{Inverse participation ratio of all the eigenstates $\psi_{k}$ for the logistic potential with $r = 4.0$. In the boxes there appear the choices used in Fig. 6.}
 \label{fig5}
  \end{center}
\end{figure}

\begin{figure}[h!]
\begin{center} 
 \includegraphics[height=7cm, width=12cm]{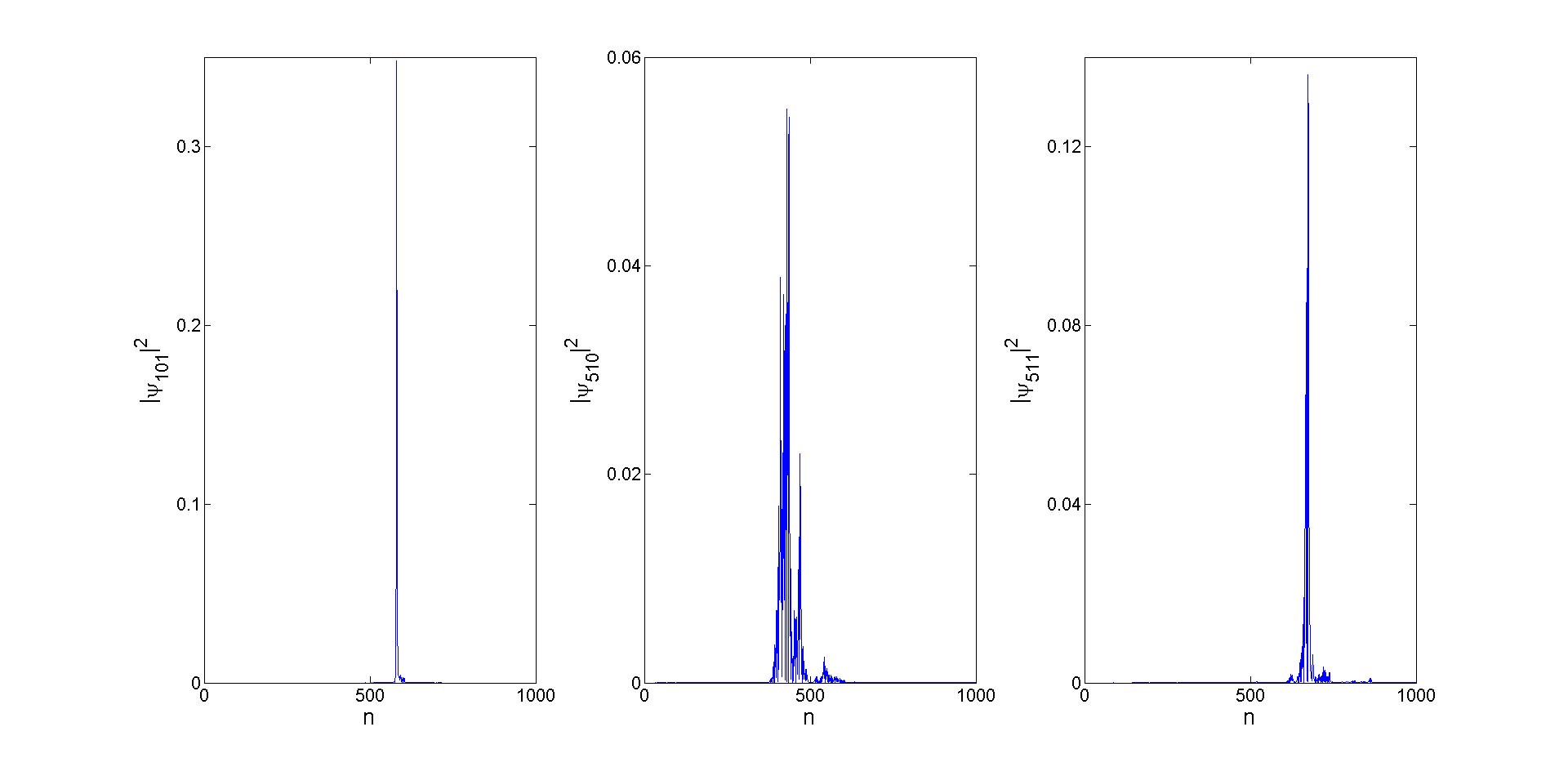} 
 \caption{Logistic Map. Probability density distributions over the lattice sites $n$ for the eigenstates $\psi_{101}$, $\psi_{510}$, and $\psi_{511}$.}
 \label{fig6}
  \end{center}
\end{figure}

In this work, chaotic maps were intentionally chosen due to their intricate dynamics, which provided the desired disorder, and to the positive Lyapunov exponents in chaotic regions, which provided the desired control. Nonetheless, there is a whole class of systems, so-called strange nonchaotic attractors (SNA) \cite{gopy1984}, which present a fractal nature, with complicated dynamics, but for which all the Lyapunov exponents are either zero or negative. It would be interesting to test the behavior of the participation ratio $IPR$ for tight-binding solutions generated by binary sequences built from such attractors. In what follows, as an example, we take the quasiperiodically forced logistic map \cite{heagy1994,prasad1997,prasad2001} given by
\begin{eqnarray}
{w_{n + 1}}& = & r\left[ {1 + \epsilon \cos \left( {2\pi {\theta _n}} \right)} \right]{w_n}\left( {1 - {w_n}} \right)\\
{\theta _{n + 1}} & = & {\theta _n} + \phi \;\;\; \left( {\bmod 1} \right),
\end{eqnarray}
where $\phi$ is an irrational number (we take the golden mean $\phi = \left(\sqrt{5} - 1 \right)/2$), and the usual restriction $0 \le r \le 4$ imposes $0 \le \epsilon \le (4/r - 1)$. It was shown that for $\epsilon ' = r\epsilon /\left( {4 - r} \right)= 0.95$ this system has a SNA behavior in the approximate interval $2.75 \le r \le 3$. For $r$ far below this interval, one has (quasi)periodic dynamics, which becomes more and more complicated as $r$ approaches the SNA region. For $r > 3$ one has two regions of chaotic behavior separated by a narrow (quasi)periodic region around $r \approx 3.4$. In Fig. 7 we can see that the inverse participation ratio $IPR(r)$ signals all these features. We take as an example a center-band delocalized eigenstate, $\psi_{300}$.  $IPR(r)$ presents small values for $r$ in the periodic region near $r = 2$ and attains increasing values as $r$ approaches the SNA region and the dynamics becomes more complicated. Even in the SNA region, where the  (largest) Lyapunov exponent $Ly \left( r \right) = \frac{1}{N}\sum\nolimits_{n = 1}^N {\ln \left| {r\left[ {1 + \epsilon \cos \left( {2\pi {\theta _n}} \right)} \right]\left( {1 - 2w_n } \right)} \right|}$ oscilates slightly around zero, $IPR(r)$ correctly points the intricateness of the dynamics, with a corresponding localization process of the eigenstate (not shown here). It could be interesting to note that this results corroborates in a sense the choice of chaotic maps as a recipe for controled disorder since, then, the vanishing $IPR$ in periodic regions clearly distinguishes it from the chaos-given disordered behavior with $IPR(r) > 0$.

\begin{figure}[h!]
\begin{center} 
 \includegraphics[height=7cm, width=12cm]{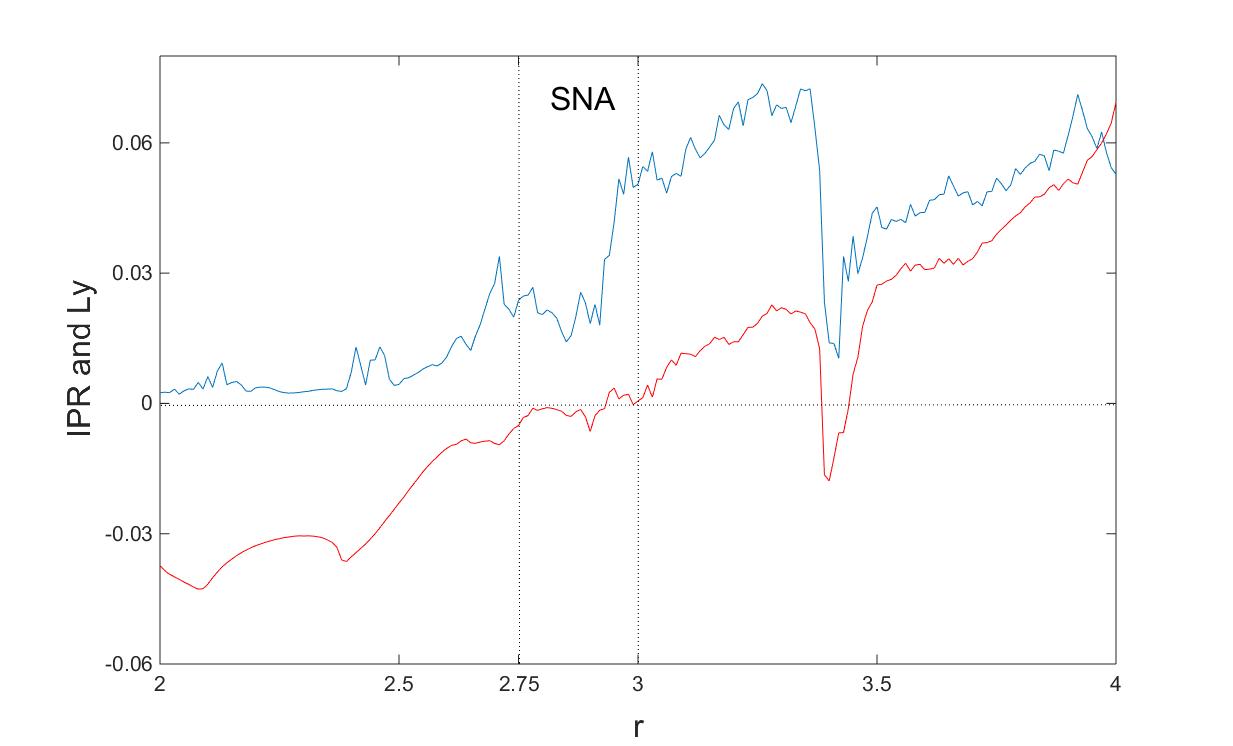} 
 \caption{Forced Logistic Map. $IPR(r)$ and $Ly(r)$ multiplied by $0.1$ for better viewing. Dotted guiding lines mark zero for $IPR$ and $Ly$, and the approximate SNA region.}
 \label{fig7}
  \end{center}
\end{figure}

Finally, we just mention that, although it is known that the strength of the potential influences the localization properties of the eigenstates, the inverse participation ratio consistently keeps the agreement seen in the cases presented here. We have checked this consistency for different values of the potential amplitude $V_n$ in the interval $\left[ 0,1 \right]$.

\section{Conclusion}\label{sec:6}

In this work we have proposed the use of new binary sequences, based on three different chaotic maps, as a way of introducing variable-disordered potentials in the Schr\"odinger operator. We have seen that the large-$N$ value of the Inverse Participation Ratio ($IPR(r)$) for the ground-state of the Schr\"odinger operator, as a function of the chaoticity parameter $r$, correctly indicates the windows of periodicity and the regions of chaos pointed out by the Lyapunov exponents $Ly(r)$. Moreover, we have seen in figures 1--3 (b) that $IPR(r)$ follows the increasing (decreasing) behavior of $Ly(r)$, providing a measure of chaoticity and disorder for the binary sequences, where the Lyapunov exponent is not directly applicable.

Finally, we mention that the techniques of growing superlattices by epitaxy \cite{axel1991} could in principle be applicable to the chaotic potentials considered in this work. Experimental, and eventually technological, exploration of systems based on these chaotic potentials could then be considered.\\

\section*{Acknowledgements}
The authors thank sincerely Luciano Coutinho dos Santos and Luiz Argel Poveda Calvino for discussions on nonchaotic attractors. Also, the authors thank very much the anonymous Referees for their invaluable advice.

\bibliography{gianbibfile}
\end{document}